\documentclass{ws-mpla-mod}
\usepackage[super]{cite}
\usepackage{graphicx}

\usepackage{color,soul} 
\usepackage{xcolor} 

\begin{document}

\markboth{Mohamed Lamine Abdelali, Noureddine Mebarki, Bouthaina Boutaghane and Maimana Trirat}{New MiniBooNE excess interpretation using Pseudo-Hermitian neutrino oscillation}

\catchline{}{}{}{}{}

\title{New MiniBooNE excess interpretation using Pseudo-Hermitian neutrino oscillation}

\author{Mohamed Lamine Abdelali}
\address{Sciences Departement, Teacher Education College of Setif, P B. 556  El Eulma,\\ Setif, 19600, Algeria\\
Laboratoire de Physique Mathematique et Subatomique, Universite Freres Mentouri Constantine 1, BP 325 Route de Ain El Bey, \\ Constantine, 25017, Algeria\\
m.abdelali@ens-setif.dz,\\ mabdelali1@gmail.com\\}

\author{Noureddine Mebarki}
\address{Laboratoire de Physique Mathematique et Subatomique, Universite Freres Mentouri Constantine 1, BP 325 Route de Ain El Bey, \\ Constantine, 25017, Algeria\\}

\author{Bouthaina Boutaghane}
\address{Sciences Departement, Teacher Education College of Setif, P B. 556  El Eulma,\\ Setif, 19600, Algeria\\}

\author{Maimana Trirat}
\address{Sciences Departement, Teacher Education College of Setif, P B. 556  El Eulma,\\ Setif, 19600, Algeria\\}

\maketitle

\pub{Received (Day Month Year)}{Revised (Day Month Year)}

\begin{abstract}
The excess observed in the MiniBooNE experiment data has been confirmed by several data releases and analyses. 
Several attempts to explain this excess have been proposed in the literature, using approaches ranging from experimental errors to new neutrino theories. 
None of these attempts seem to fully account for the observed excess. 
In this paper, we present a new approach based on a non-standard two-flavor oscillation theory. 
In recent years, there has been increasing interest in pseudo-Hermitian Hamiltonian theories and their impact on various quantum phenomena. 
Neutrino oscillation is modified using the approach of PT-pseudo-Hermitian oscillation, which introduces a zero-distance effect and an asymmetry in the appearance probability of specific flavors. 
According to this study, such effects can account for the excess and provide a better fit to the data than the current best fit within the standard oscillation framework. 
We present an analysis of the fit between MiniBooNE data and this new oscillation model. 

\end{abstract}

\ccode{PACS Nos.: include PACS Nos.} 

\section{Introduction}
\label{s.0}
The MiniBooNE experiment Ref.~\refcite{Ref.00.01}  found a significant excess of low-energy {$\nu _{e}$}-like events. 
A few years later, an updated study Ref.~\refcite{Ref.00.02} by the MiniBooNE collaboration confirmed the observation of a total excess of approximately 638 electron-like events in the energy range between 200 MeV and 1250 MeV in both neutrino and antineutrino running modes. 
The new study examined the neutrino oscillation results using increased data and updated background studies. 
However, the overall significance of the excess still ranges from $4.8\sigma$, limited by systematic uncertainties, up to $12.2\sigma$ for the statistical significance, as concluded in Ref.~\refcite{Ref.00.02}. 
Ref.~\refcite{Ref.Ex.001} provides a summary of all MiniBooNE data releases, which can be accessed on a public website \footnote{MiniBooNE data release website: https://rtayloe.pages.iu.edu/MB/data-releases/}.

{The  MiniBooNE excess} is in conflict with {currently accepted results of neutrino research} as it does not fit with the data of other short-baseline neutrino oscillation experiments Ref.~\refcite{Ref.00.03}, Ref.~\refcite{Ref.00.04} and Ref.~\refcite{Ref.00.05}. 
Several explanations of this excess have been proposed. 
A short-baseline {$\nu _{\mu}$ to $\nu _{e}$} oscillations generated by active-sterile neutrino mixing has been proposed by Ref.~\refcite{Ref.00.07}, Ref.~\refcite{Ref.00.08} and Ref.~\refcite{Ref.00.09}. 
Other physics beyond the Standard Model are proposed in Ref.~\refcite{Ref.00.10}, Ref.~\refcite{Ref.00.11}, Ref.~\refcite{Ref.00.12} and Ref.~\refcite{Ref.00.13}. 
Some of the proposed solutions explain the excess as non-neutrino events misinterpreted as neutrinos and anti-neutrinos. 
Non-oscillatory explanations are proposed but not accepted taking {into account} other measurements Ref.~\refcite{Ref.00.06}, which is also the case for all the above proposed explanations. 

 {In quantum mechanics, the Hamiltonian is typically required to be Hermitian to ensure real eigenvalues (which correspond to observable energy levels) and unitary time evolution (which preserves probability). 
However, non-Hermitian Hamiltonians may still yield physically meaningful results if they satisfy certain symmetry conditions, particularly $PT$ symmetry and pseudo-Hermiticity (see Ref.~\refcite{Ref.Ex.002}). 
If $PT$ symmetry is unbroken, the eigenvalues of $H$ can be real, even if $H$ is non-Hermitian. 
This allows such Hamiltonians to describe physically meaningful quantum systems. 
If $PT$ symmetry is spontaneously broken, eigenvalues become complex, leading to instabilities or decay. 
This transition is relevant in non-equilibrium quantum mechanics and quantum phase transitions  (see Ref.~\refcite{Ref.Ex.004} and Ref.~\refcite{Ref.Ex.005}). 
A Hamiltonian $H$ is pseudo-Hermitian if there exists a metric operator such as that $H$ can be mapped to an equivalent Hermitian Hamiltonian via a similarity transformation. 
The appropriate physical inner product is not the standard one but defined via the metric, ensuring unitary evolution. 
If $H$ is both $PT$-symmetric and pseudo-Hermitian, it often has a deeper connection to a Hermitian Hamiltonian in a modified inner product space (see Ref.~\refcite{Ref.Ex.003}). 
This guarantees real eigenvalues and a consistent probability interpretation. 
For the physical Interpretations and applications, $PT$ symmetry describes gain-loss-balanced optical waveguides, leading to real propagation constants and exceptional points in photonics and optical systems (see Ref.~\refcite{Ref.Ex.006}). 
$PT$-symmetric quantum field theories can describe phase transitions and non-Hermitian deformations of the Standard Model. 
Effective non-Hermitian Hamiltonians govern dissipation, quantum decay, and non-equilibrium steady states in open quantum systems. 
Non-Hermitian topological phases emerge in $PT$-symmetric lattices and systems with exceptional points as shown in condensed matter studies (see Ref.~\refcite{Ref.Ex.007} and Ref.~\refcite{Ref.Ex.008}).} 

 {In the context of neutrino oscillations, several studies explored the application of $PT$-symmetric Hamiltonians and their influence on neutrino flavor oscillations.
For instance, in Ref.~\refcite{Ref.Ex.009}, Ohlsson extends the conventional two-flavor neutrino oscillation framework with a novel approach by incorporating a non-Hermitian $PT$-symmetric effective Hamiltonian. 
This approach derives conditions under which the Hamiltonian maintains a real spectrum, ensuring physically meaningful predictions. 
It suggests that such extensions can introduce sub-leading effects in neutrino flavor transitions related to those observed with non-standard neutrino interactions. 
Additionally, in Ref.~\refcite{Ref.0001}, the researchers define transition probabilities between two flavor states in a quantum system described by a $PT$-symmetric non-Hermitian Hamiltonian. 
They perform explicit calculations to demonstrate probability conservation when a proper definition of the final state is adopted and applying these principles to neutrino oscillations in vacuum. 
Furthermore, in Ref.~\refcite{Ref.Ex.010}, the authors investigates the behavior of transition probabilities in systems where $PT$ symmetry is broken. 
The study's insights could be relevant for understanding neutrino oscillations under specific conditions. 
Another pertinent study by Jean Alexandre et al. formulates transition matrix elements consistent with positivity and perturbative unitarity in $PT$-symmetric non-Hermitian systems (See Ref.~\refcite{Ref.Ex.011}). 
The authors discuss potential applications to the oscillations of mesons and neutrinos, suggesting that such a framework could provide insights into observed neutrino oscillation patterns.
Authors of Ref.~\refcite{Ref.Ex.012} consider the $PT$ symmetric non-Hermitian Hamiltonian of two flavour neutrino case to probe whether neutrinos are Dirac or Majorana. 
Some experimental setups are proposed to explore the non-Hermiticity in quantum phase transitions and understand the behavior of dissipative quantum systems, implying significant development of new quantum devices and technologies (See Ref.~\refcite{Ref.Ex.013}). 
A relativistic representation has recently been presented for a class of quantum mechanical particle oscillations. 
The resulting quantum mechanical evolution is shown to be unitary and probability is conserved in the oscillations (See Ref.~\refcite{Ref.Ex.014}). 
A conprehensive summary of the recent progress in the formulation of flavor mixing and oscillations in pseudo-Hermitian quantum theories can be found in Ref.~\refcite{Ref.Ex.015}.} 

In this paper, we propose a new explanation for the MiniBooNE excess by including a new oscillation framework. 
Section 2 briefly presents the standard model of oscillation with a comparaison with the model of pseudo-hermitian oscillation. 
In section 3, we present a {new visualization} of the MiniBooNE data {to provide a better insight into the excess}. 
In section 4, we discuss our analysis of the excess in the data using the new oscillation model and its implications. 
Finally, we {provide} our conclusions and {some future} perspectives. 

\section{Neutrino oscillations models}
\label{s.1}
We briefly describe in this section the standard model of neutrino oscillations based on neutrino mass eigenstates and a new neutrino oscillations model based on a $PT$-symmetric pseudo-hermitian hamiltonian physics. 

\subsection{Standard Two Flavour Oscillations}
\label{ss.1.1}
The standard neutrino oscillation model is based on the propagation through space of massive neutrinos, which differ from the {observed flavour} neutrinos. 
The mass difference and the mixing angle between the massive neatrinos are responsible {for} the phenomenon of flavor neutrinos oscillations. 
In {this} phenomenon, the observed flavour of the neutrino changes during its propagation, depending on its energy and the {travelled distance}. 
The standard oscillations probability $P_{ab}$ between two flavours of neutrinos (${\nu}_{a}$ and ${\nu}_{b}$) {is} expressed as follows 
\begin{equation}
\label{eq.1.1}
P_{a b}= \sin ^2 (2 \theta) \sin ^2 \left[ 1.27 \left( \frac{\Delta m^2}{eV ^2} \right) 
\left( \frac{L_{\nu}}{km} \right) \left( \frac{GeV}{E_{\nu}} \right)
  \right] 
\end{equation}
where: $ \theta$ is the mixing angle between the two flavors, $\Delta m^2$ is the squared mass difference between the two massive neutrinos expressed in eV, $L$ is the {traveled distance} by neutrinos from creation to observation expressed in km {and}  $E$ is the energy of the neutrino expressed in GeV. 
This is called appearance probability, which describes the probability to observe a specific neutrino flavour from an initial beam of a different flavour. 

In the standard model of neutrino oscillations, the appearance probabilities of the two flavours are the same due the hermicity of the Hamiltonian 
\begin{equation}
\label{eq.1.2}
P_{b a} = P_{a b}
\end{equation}
Then, due the unity of the sum of all probabilities for {a specific flavour}, we could express the disappearance probability as
\begin{equation}
\label{eq.1.3}
P_{a a} = P_{b  b} = 1- P_{a b}
\end{equation}
which describes, in a statistical perspective, the percentage of {observed neutrinos in their intial flavour}. 

\subsection{Pseudo-Hermitian Hamiltonian}
\label{ss.1.2}
The $PT$-symmetric pseudo hermitian Hamiltonian is studied in Ref.~\refcite{Ref.0001} . 
In this section, we present a brief review of their results and an interpretation of the most important predictions. 

The Hamiltonian of this model has the following form 
\begin{equation}
\label{eq.1.4}
H= 
\begin{pmatrix}
\rho e^{i \varphi} & \sigma \\
\sigma          & \rho e^{-i \varphi}
\end{pmatrix}
\end{equation}
where $\rho$, $\varphi$ and $\sigma$ are {Pseudo-Hermitian Hamiltonian parameters}. 
In this framework, the oscillation probabilities between two neutrinos, ${\nu}_{a}$ and ${\nu}_{b}$, are different from the standard model of neutrinos oscillations. 
The appearance probabilities are given as follows
\begin{equation}
\label{eq.1.5}
P_{a b}= \sin ^2 \left( \frac{\alpha}{2} - \frac{\beta}{2} t \right)
\,\, , \,\,
P_{b a}= \sin ^2 \left( \frac{\alpha}{2} + \frac{\beta}{2} t \right)
\end{equation}
The disappearance probabilities are given by 
\begin{equation}
\label{eq.1.6}
P_{a a}= \cos ^2 \left( \frac{\alpha}{2} - \frac{\beta}{2} t \right)
\,\, , \,\,
P_{b b}= \cos ^2 \left( \frac{\alpha}{2} + \frac{\beta}{2} t \right)
\end{equation}
where
$
\sin \alpha= \frac{\rho \sin \varphi}{\sigma}
$
   ,   
$
\beta = 2 \sqrt{ \sigma ^2 - \rho ^2 \sin ^2 \varphi} = 2 |\sigma \cos \alpha|
$. 
It is also possible to write the new hamiltonian in term of neutrino energy as
\begin{equation}
\label{eq.1.7}
H= \frac{1}{4 E}
\begin{pmatrix}
\rho e^{i \varphi} & \sigma \\
\sigma          & \rho e^{-i \varphi}
\end{pmatrix}
\end{equation}
which modifies the probability to be 
\begin{equation}
\label{eq.1.8}
P_{a b}= \sin ^2 \left[ \frac{\alpha}{2} - 1.27 |\cos \alpha| \left( \frac{|\sigma|}{eV^2} \right) \left( \frac{GeV}{E} \right) \left( \frac{L}{km}\right) \right]
\end{equation}
 {The authors of Ref.~\refcite{Ref.0001} applied this formalism and oscillation probability to two-flavor neutrino oscillations of $\nu _{\mu}$ to $\nu _{\tau}$ and $\nu _{\mu}$ to $\nu _{\mu}$ in vacuum. 
It is important to note that The hamiltonian of Eq.~(\ref{eq.1.7}) and  the oscillation probability of Eq.~(\ref{eq.1.8}) are simplified forms of Eq.~(3.2) and Eq.~(3.10) in Ref.~\refcite{Ref.0001} as it considers no mass or mass difference between the two flavor states. 
Then, in the following analysis, the oscillations between the two neutrino flavors are due solely to $PT$-symmetric hermitian property of the hamiltonian with no need for a mass eigenstates. 
Such a hypothesis, if confirmed by data from MiniBooNE and other collaborations would change our perspective on neutrino oscillations, especially as it does not violate the massless neutrino hypothesis. 
Moreover, the impact of the mass difference cannot be distinguished from the contribution of $\sigma$ as the probability of Eq.~(3.10) is dependent on their sum.} 

By comparing the new oscillation probabilities to the standard oscillation model, the most important two aspects of {these oscillations are} 
\begin{romanlist}[(b)]
\item \textbf{zero-distance effect}, which is related to $\alpha$ value, which create a sort of a phase. 
Then, even without propagation, the neutrino observed at its source has a non-zero oscillation probability. 
\item \textbf{Asymmetry of the probability}, observed as the difference between $P_{a b}$ and $P_{b a}$ as well as between $P_{a a }$ and $P_{b b}$. 
This difference could enhance the oscillation toward a specific flavor depending on distance traveled. 
\end{romanlist}
To {test the} validity of such a model, experimental data should be fitted to the model and analysed by using these new probabilities for the theoretical predictions. 

\section{Peculiar results of MiniBooNE data }

MiniBooNE is a short-baseline detector aiming to observe the appearance of $\nu _e$ from the oscillation of $\nu _{\mu}$ in a well constrained background and also of $\bar{\nu} _e$ from $\bar{\nu} _{\mu}$. 
The fit to a standard two-flavor oscillations is not good, since even the best fit leaves part of the excess {between theoretical predictions and experimental measurements}. 
Moreover, the best fit is in conflict with results from other collaborations. 
 {The best combined neutrino oscillation fit occurs at ($\sin^2 (2\theta)$, $\Delta m^2$) = (0.92, 0,041 eV$^2$), as well as at a point within 1$\sigma$ of the best-fit point (0.01, 0,4 eV$^2$). 
And the different confidence contours of MiniBooNE do not match exactly the confidence level limits from the KARMEN and OPERA experiments as shown in Ref.~\refcite{Ref.00.02} plots.} 
The available data release of 2018 described in Ref.~\refcite{Ref.00.01} and Ref.~\refcite{Ref.00.02} consist of predicted and observed candidate events in each energy bin from 200 MeV to 3 GeV for $\nu _e$ and $\nu _{\mu}$ in neutrino mode and for $\bar{\nu} _e$ and $\bar{\nu} _{\mu}$ in antineutrino mode. 
Contour points are also provided around allowed regions of 1 $\sigma$, 90\% and 99\% confidence level (CL) in the ($\sin^2 2\theta$ , $\Delta m^2$) plan. 

We present a different {visualization} of the data to {better understand its peculiarities}. 
This representation differ from those present in Ref.~\refcite{Ref.00.01} and Ref.~\refcite{Ref.00.02}. 
We plot the ratio $R_{Observed/Expected}$, defined as the number of observed events over the number of predicted events of a specific neutrino or anti-neutrino flavor in each energy bin. 
An excess of events occurs when the number of observed events exceeds the theoretical expectation. 
It is equivalent to a ratio greater than one in our plots. 
The deficit occurs when the number of observed events is less than the theoretical expectation. 
This is shown in our plots as a ration less than one. 

\begin{figure}[tbp]
\centerline{
\includegraphics[width=7cm,height=6cm]{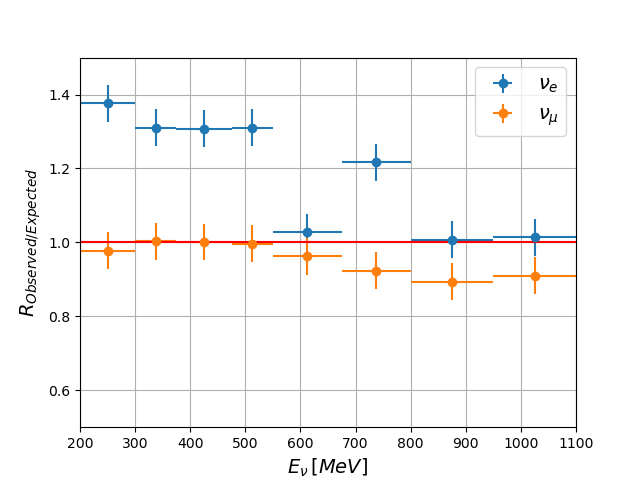} 
\includegraphics[width=7cm,height=6cm]{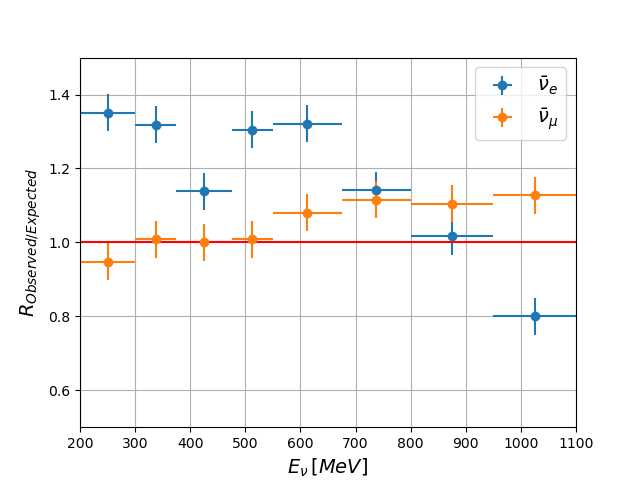} 
}
\vspace*{8pt}
\caption{\protect\label{fig.2.1} Reproduction of the observed-to-predicted ratio of neutrino events from the 2018 MiniBooNE data release. Neutrino flavor mode in the left side and antineutrino flavor mode in the right side. }
\end{figure}

in Fig.~\ref{fig.2.1} (left side), We observe varying excesses in the electron neutrino events accoss different bins. 
Conversely, deficits are observed in muon neutrino events across the bins. 
The excesses and deficits are not uncorrelated across the energy bins. 
For antineutrinos, a different pattern of excesses and deficits appears for the two favours, as shown in Fig.~\ref{fig.2.1} (right side). 
In some bins, both muon and electron antineutrino exhibit excess. 
The patern of excess and deficit reverses at higher energies, where electron antineutrinos show excess and the muon antineutrinos show deficit, switching again at highier energies. 

\begin{figure}[tbp]
\centerline{ 
\includegraphics[width=7.5cm,height=6.5cm]{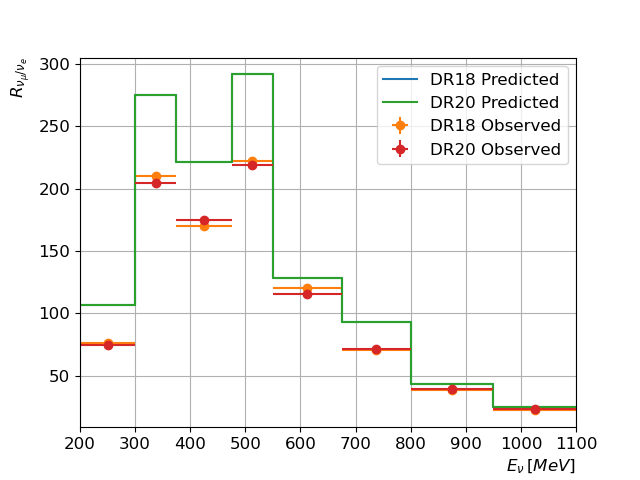} 
\hfill
\includegraphics[width=7.5cm,height=6.5cm]{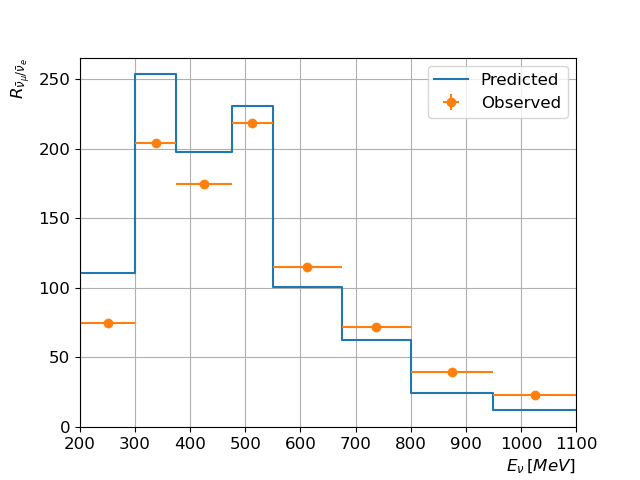} 
}
\vspace*{8pt}
\caption{\protect\label{fig.2.3} Reproduction of the flavour ratio for observed and predicted events. The left side show neutrino mode for both data releases of 2018 and 2020. The right side show antineutrino mode for data release of 2020.}
\end{figure}

 {Another representation is the flavor ratio $R_{\alpha/\beta}$ of $\nu _{\mu}$ ($\bar{\nu} _{\mu}$) to $\nu _{e}$ ($\bar{\nu} _ {e}$) for observed (final) data and for predicted (initial) background without oscillations.} 
In this representation, for the neutrino mode, Fig.~\ref{fig.2.3} (left side) shows a tendecy toward a higher electron composition in the final beam at the detector across all energy bins. 
In the antineutrino mode, an excess appears in some bins and a deficit in others in electon antineutrino conposition of the final beam, as {shown} in Fig.~\ref{fig.2.3} (right side). 
 {A $\chi ^2$ goodness-of-fit test comparing predicted to observed $R_{\nu _e/\nu _{\mu}}$  shows no significant difference and cannot exclude the null hypothesis. 
This is primarely due to the big difference between the event numbers as the muon event numbers are much higher. 
However, performing the same $\chi ^2$ test with the the ratio reversed to be $R_{\nu _{\mu}/\nu _e}$ reveals a significance difference, allowing the full data patterns to be represented statistically sgnificantly. 
Form this point, we use $R_{\nu _{\mu}/\nu _e}$ and $R_{\bar{\nu} _{\mu}/\bar{\nu} _e}$ in the following analysis, abbriviated as $R_{\alpha / \beta}$.} 

\section{New MiniBooNE data analysis}

In light of the previous sections, it is clear the an inconsistency between the observed number of events and those predicted by the current {standard} oscillation model. 
Diffenrent studies, including the MiniBooNE collaboration's analysis, failed to explain this excess as experiment errors {or to identify the cause of these measurements}. 
The posibility of non-oscillation explanations or extra events contributing to the observed excess has also been ruled out. 
{Thus, the most adequate explanation is the existence of} a new model of neutrino oscillations that could be more consistent with the data. 
This new oscillation model is related to new physics applied to neutrinos, either generally or specifically in low energy, {short} distances, or both. 
The model used in this paper, presented in section~\ref{ss.1.2}, is based on $PT$-symetric pseudo-hermitian oscillations. 
 {We use the MiniBooNE data from 2018 and 2020.} 

To revisit MiniBooNE data, a new method should be uadopted for the following reasons  

\begin{romanlist}[(b)]
\item Limitations of released data, as we do not have detailed expected events from each possible sources, but a total per bin, 
\item Energy bins of $\nu  _{\mu}$ differ from those of $\nu _e$, 
\item In the case of pseudo-Hermitian oscillations, the asymmetry of probabilities should be fully accounted for, 
\item There is no access to the systematic uncertainties related to the flux and detector.
\end{romanlist}

In our analysis, we compare  {flavour event ratios} rather than event number directly. 
 {In flavour event ratios}, differences in oscillation probabilities can be more clearly observed. 
Firstly, the final number $N^f_{\alpha}$ of a specific neutrino ${\nu}_{\alpha}$ is calculated from initial numbers ($N^i_{\alpha}$, $N^i_{\beta}$) and can be expressed as 
\begin{equation}
\label{eq.3.1}
N^f_{\alpha} =  N^i_{\alpha} P_{\alpha \alpha} + N^i_{\beta} P_{\beta \alpha}
\,\, , \,\,
N^f_{\beta} =  N^i_{\beta} P_{\beta \beta} + N^i_{\alpha} P_{\alpha \beta}
\end{equation}
where $P_{\alpha \beta}$ is the oscillation probability. 
Secondly, the final ratio of the two flavours at the detector are given by
\begin{equation}
\label{eq.3.2}
\frac{N^f_{\alpha}}{N^f_{\beta}}  = \frac{N^i_{\alpha} P_{\alpha \alpha} + N^i_{\beta} P_{\beta \alpha}}
{N^i_{\beta} P_{\beta \beta} + N^i_{\alpha} P_{\alpha \beta}}
\end{equation}
This ration can be further simplified to obtain 
\begin{equation}
\label{eq.3.3}
\frac{N^f_{\alpha}}{N^f_{\beta}}  = \frac{\frac{N^i_{\alpha}}{N^i_{\beta}} P_{\alpha \alpha} + P_{\beta \alpha}}
{P_{\beta \beta} + \frac{N^i_{\alpha}}{N^i_{\beta}} P_{\alpha \beta}}
\end{equation}
Finally, the final ratio $R^f_{\alpha / \beta}$ between the number of neutrinos of two flavours is given by
\begin{equation}
\label{eq.3.4}
R^f_{\alpha / \beta}  = \frac{R^i_{\alpha / \beta} P_{\alpha \alpha} + P_{\beta \alpha}}
{P_{\beta \beta} + R^i_{\alpha / \beta} P_{\alpha \beta}}
\end{equation}
This final ratio $R^f_{\alpha / \beta}$ depends on the initial ratio $R^i_{\alpha / \beta}$ at the source and all oscillation probabilities ($P_{\alpha \alpha}$, $P_{\beta \alpha}$, $P_{\beta \beta}$, $P_{\alpha \beta}$). 
These probabilities are calculated using a specific model. 
This method is applied to both pseudo-hermitian oscillation and standard oscillations  {to compare their predictions and how well they fit the MiniBooNE collaboration data.} 
The initial  {flavour} ratio is calculated from the predicted event numbers without oscillation, as provided in the released MiniBooNE data. 

To test the model's fit to the data, we calculate $\chi ^2$ parameter using the final theoretical and observed ratios. 
A lower $\chi ^2$ value  indicates a better fit of the model, with specific parameters, to the data. 
{The definition of the $\chi ^2$ parameter is given by} 

\begin{equation}
\label{eq.3.6}
\chi ^2 = \frac{ 
\left( R^{Model}_{\alpha / \beta} - R^{Data}_{\alpha / \beta} \right) ^2
}{
R^{Model}_{\alpha / \beta}
}
\end{equation}

{To obtain the best fit for a specific model, we scan its parameter space by creating a mesh of combinaisions of the two model parameters. 
The parameter values are incremented by $0.001$ in the range from $0.0$ to $1.0$. 
We calculate the $\chi ^2$ value for each conbination of the model's free parameters. 
The set of combinations represents a test over all  {possible values of parameters space}. 
In the standard oscillation model, the free parameters are $\theta$ and $\Delta m^2$, which have been estimated by various collaborations using their experimental data. 
For the pseudo-hermitian model, the free parameters are $\alpha$ and $\sigma$, which will be determined using MiniBooNE data. 
Those parameters influence the oscillation probabilities. 

\begin{figure}[tbp]
\centerline{
\includegraphics[width=7.5cm,height=6.5cm]{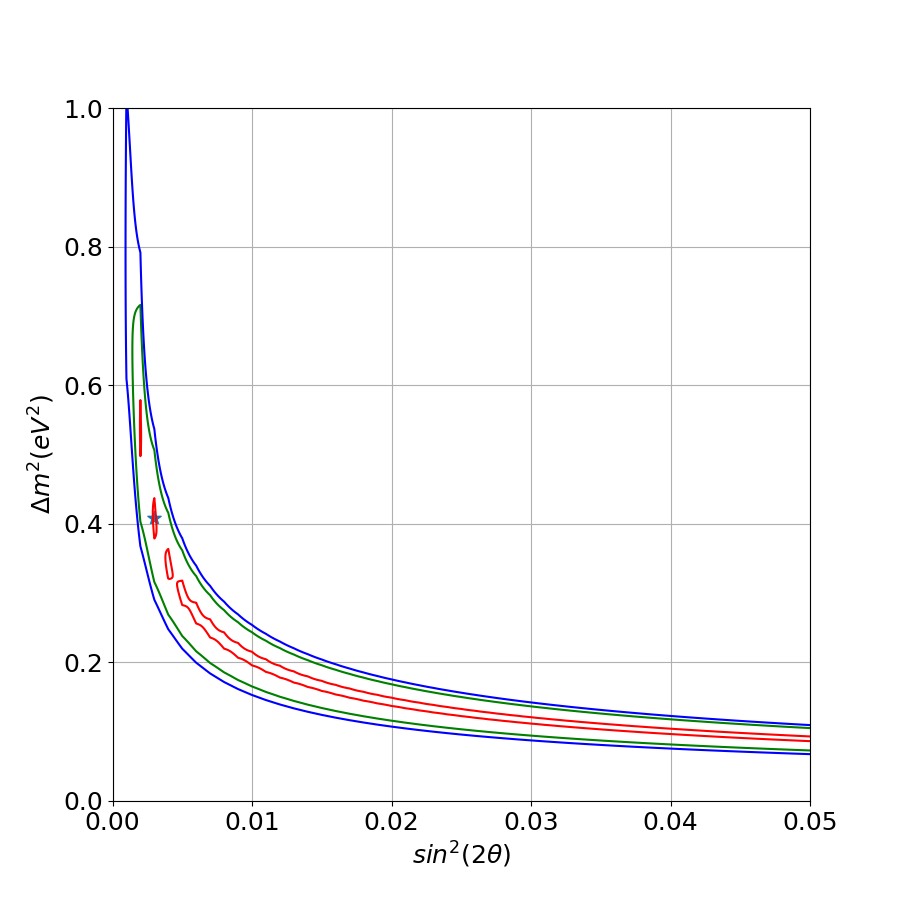}
\hfill
\includegraphics[width=7.5cm,height=6.5cm]{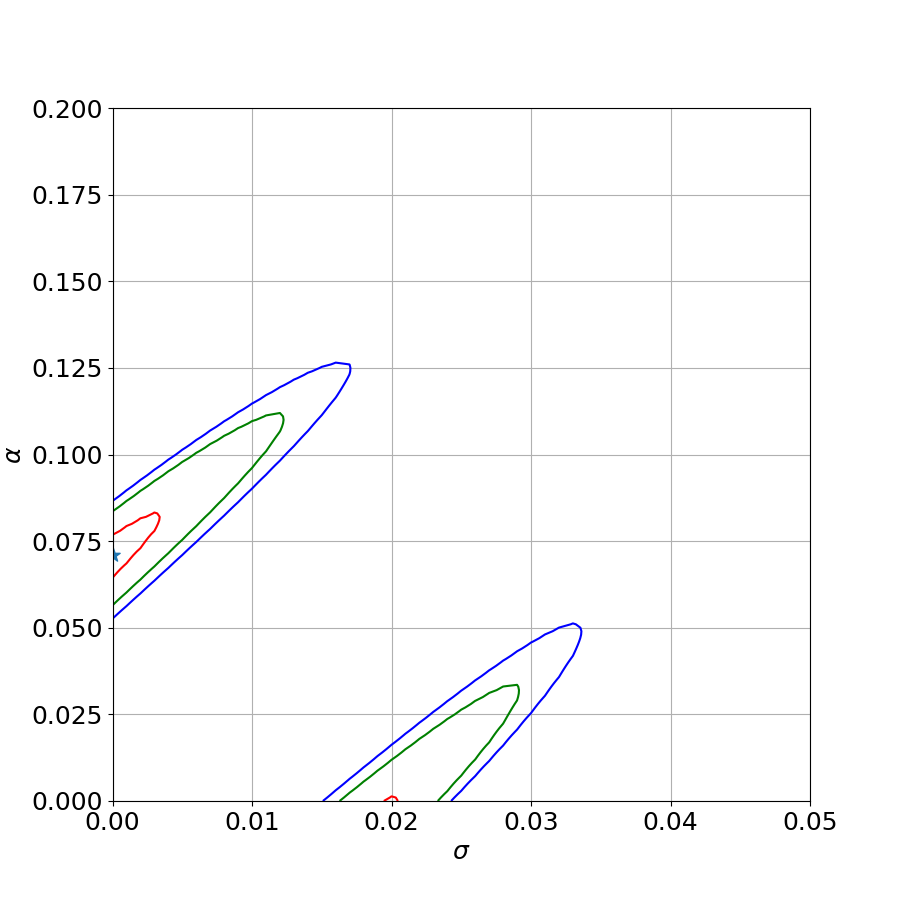}
}
\vspace*{8pt}
\caption{\protect\label{fig.3.2} Contour lines and the best fit MiniBooNE data release of 2020 for standard oscillation model (left side) and pseudo-hermitian hamiltonian model (right side) using combined $\chi^2$ estimation for both neutrinos and antineutrinos.}
\end{figure}

\begin{figure}[tbp]
\centerline{
\includegraphics[width=7.5cm,height=6.5cm]{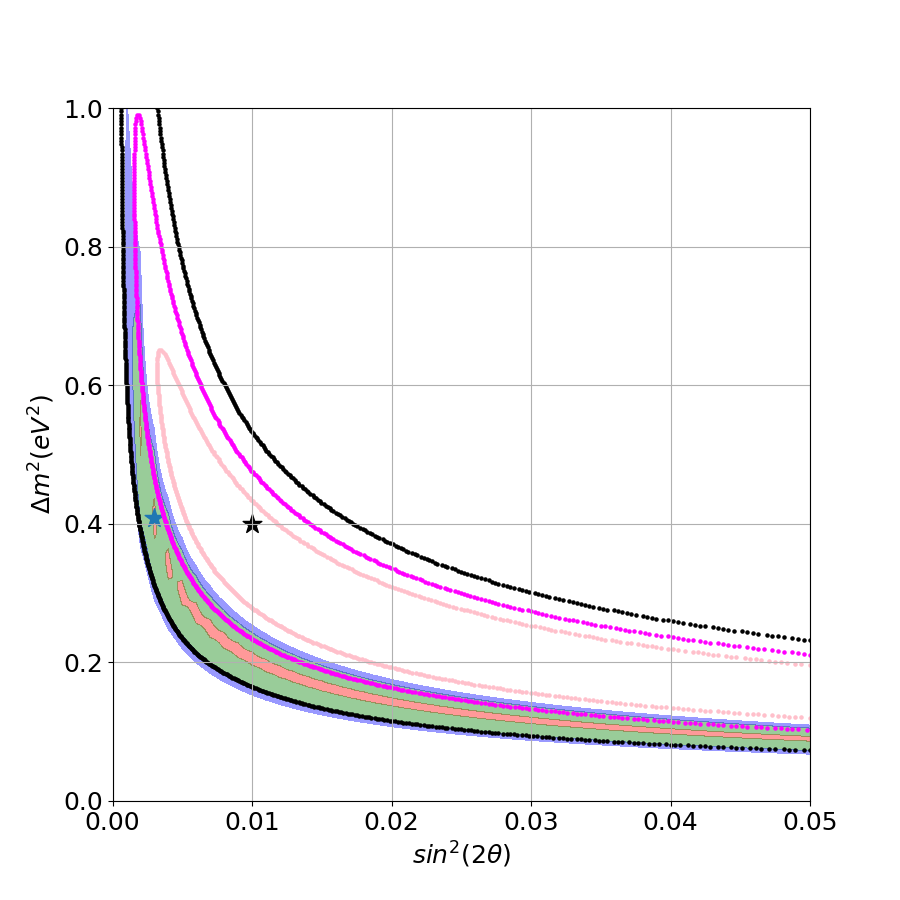}
}
\vspace*{8pt}
\caption{\protect\label{fig.3.12} Comparison between ($68\%$,$95\%$,$99\%$) Confidence level contours from MiniBooNE analysis of 2018 and the analysis of this study. Red, green and blue regions represent the contours of this analysis. The dotted lines in pink, magenta and black represent the MiniBooNE analysis. The best fit is represented by the star within the confedence regions. }
\end{figure}

\begin{figure}[tbp]
\centerline{
\includegraphics[width=7.5cm,height=6.5cm]{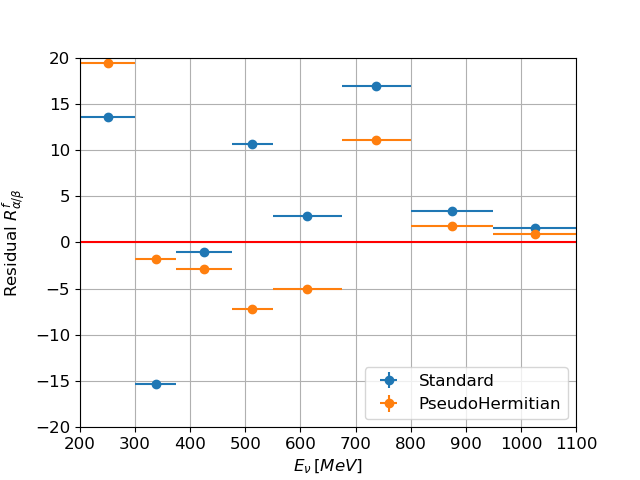}
\hfill
\includegraphics[width=7.5cm,height=6.5cm]{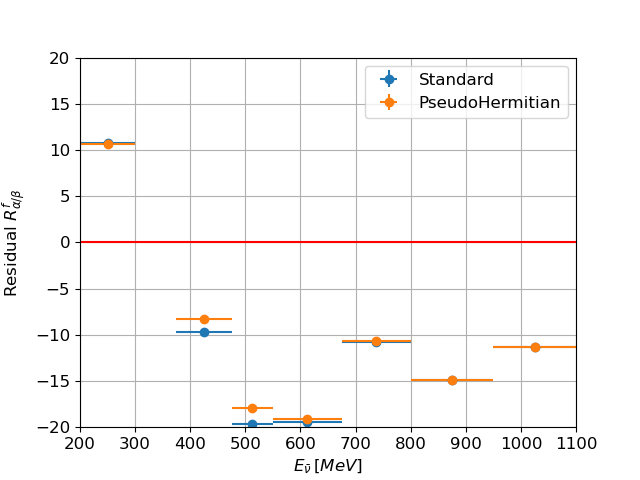}
}
\vspace*{8pt}
\caption{\protect\label{fig.3.11} Residual of observed flavour ratios between MiniBooNE data release of 2020 and the calculated ratios of the best fitting models parameters of both standard oscillations and pseudo-hermitian oscillation in the neutrino mode (left side) and antineutrino mode (right side).}
\end{figure}

 {The analysis begins with the computation of all $\chi^2$ values for each parameter combination. 
The best fit is identified as the combination yielding the lowest $\chi^2$ value. 
The best fit is determined seperately for neutrino mode, antineutrino mode, and for a combined case, where the $\chi^2$ values from both modes are summed. 
The combined case involves a larger number of data points, leading to higher critical $\chi^2$ values for the $68\%$, $95\%$ and $99\%$ confidence levels. 
The analysis is performed using a Python script available in GitHub repository \footnote{GitHub repository for the code: https://github.com/mlabdelali/MiniBoNE-Excess-Analysis}. 
The Tab~\ref{tab1} summaries the results obtained for all cases and for both MiniBooNE data releases used in this study. 
These results indicate that the best fit obtained from the new model yields a lower $\chi^2$ in both neutrino and combined modes. 
The only case where standard oscillations provides a lower $\chi^2$ is the antineutrino mode. 
It is also observed that the best-fit $chi^2$ values are consistently lower for the 2020 data release.} 

\begin{table}[h]
\tbl{\label{tab1} {The best fit for the standard oscillation (SO) model and the pseudo-Hermitian Hamiltonian (PHHO). 
The represented cases  are the fitting of the neutrino only, antineutrino only and combined between the two. 
The latest MiniBooNE data releases from 2018 (DR18) and 2020 (DR20) are used. 
The critical $\chi^2$ values are 8.145, 14.067, 18.475 for the case of neutrino-only and antineutrino-only and 16.981, 24.996 and 30.578 for the combined case.}}
{\begin{tabular}{@{}cccccccc@{}} \toprule

SO            & Best                            & Fit                     & PHHO        & Best         & Fit             & Case & Data \\
$\chi^2$ & $\sin ^2 (2\theta)$ & $\Delta m^2$ & $\chi^2$ & $\alpha$ & $\sigma$ &          &            \\
8.828 & 0.002 & 0.540 & 7.250 & 0.070 & 0.000 & Nu Only & DR18 \\
8.183 & 0.011 & 0.189 & 8.222 & 0.000 & 0.019 & AntiNu Only & DR18 \\
17.414 & 0.003 & 0.400 & 15.855 & 0.070 & 0.000 & Combined & DR18 \\
7.536 & 0.002 & 0.559 & 6.147 & 0.072 & 0.000 & Nu Only & DR20 \\
8.183 & 0.011 & 0.189 & 8.222 & 0.000 & 0.019 & AntiNu Only & DR20 \\
16.121 & 0.003 & 0.408 & 14.797 & 0.071 & 0.000 & Combined & DR20 \\
 \botrule
\end{tabular}\label{ta1} }
\end{table}

In Fig.~\ref{fig.3.2}, we present the results of the best-fit estimation for both standard and pseudo-hermitian models using $\chi^2$ minimization. 
Each plot displays closed contour regions corresponding to the three confidence levels $68\%$, $95\%$ and $99\%$. 
For the standard oscillation model, the allowed region appears as continous band in the $\sin^2 (2\theta)$, $\Delta m^2$ parameter space. 
In Ccntrast, the allowed regions for the pseudo-hermitian hamiltonian model are split into two distinct zones, centered around either null $\sigma$ or null $\alpha$. 
Interestingly, the pseudo-Hermitian model also yields $68\%$ confidence regions with non-zero values for both $\sigma$ and $\alpha$, highlighting unique properties of this model. 
For instance, a non-zero $\alpha$ introduces the so-called zero-distance effect, manifesting as an oscillation phase  even without propagation. 
Additionally, the permitted values for $\alpha$ and $\sigma$ remains low, consistent with the theoretical upper limits discussed in Ref.~\refcite{Ref.0001} which is around $10^{-3}$ for both parameters to perserve the $PT$-symmetry. 

In Fig.~\ref{fig.3.12}, we compare of the allowed regions for the standard model obtained from our analysis with those published by the MiniBooNE collaboration. 
We find significant overlap, especially within the $99\%$ confidence level region, which also contains the best-fit point of the new oscillation model. 
In Fig.~\ref{fig.3.11}, we present a comparative analysis of the best-fit parameters for both models in neutrino and antineutrino modes. 
This comparison uses residual plots, showing the differences between observed and predicted flavour ratios, to highlight how each model captures the data trends.
Fig.~\ref{fig.3.11} shows that in most bins the pseudo-hermitian oscillation model represent the closes to zero which is consistent with the lower $\chi^2$ values. 
The observations made for neutrinos are also correct for anti-neutrinos as the pseudo-hermitian oscillation model show a better fit to the observed data even if it does not account for all the observed excess  {when compared to} the standard model. 
Those results show a possible  {improvement of the explanation of the excess using} the pseudo-hermitian physics in its description of the neutrino oscillations. 

\section{Conclusions and Perspective}
The MiniBooNE data and the excess observed in the electron neutrino events poses a challenge to the current standard model of neutrino oscillations. 
Despite the update in the analysis of the collaboration study of the data, the excess is established and remains a well-established and unresolved phenomenon. 
Numerous studies has attempted to provide an explanation to this anomaly but most have fallen short in fully accounting for the magnitude of the observed excess. 
The proposed solutions range from experimental misinterpretations to theoretical extensions beyond the standard model. 

Interest in new perspectives to the quantum phenomena has grown in recent years. 
An example of these new trend is the pseudo-hermitian quantum physics. 
The $PT$-symetric pseudo-hermitian hamiltonian theorized in the litterature has provided a new model of the neutrino oscillations. 
This new model predicts peculiar properties of the oscillations like the zero-distance effect and the asymetry in the oscillation probabilities. 

We present in this paper a new analysis of the MiniBooNE data using the pseudo-hermitian oscillations as a theoritecal framework. 
The main observations from these preliminary results can be summarized as follows 
\begin{romanlist}[(b)]
\item There is no conflict between the allowed regions from the collaboration's analysis and this new analysis, 
\item The best fit for standard oscillations for ($\sin^2 2 \theta$, $\Delta m^2$) is different from the result of the collaboration, but still in their $99\%$ allowed region, 
\item $\chi ^2$ for the best fit of pseudo-Hermitian oscillations is smaller across all modes and data releases compared to standard oscillations, which make pseudo Hermitian oscillations a better fitting model, 
\item  {The confidence levels regions for the pseudo-hermitian oscillation model allow non-zero values for both model paramertes, consistent with known bounds,}
\item The excess remains present even in the best fit of the new model. 
\end{romanlist}

This study has investigated the  {possible impact} of pseudo-Hermitian physics in the case of neutrino oscillations. 
The presented results do not rule out a possible implication of this new model in the explanation of the MiniBooNE data. 
But, a more complete model should be constructed to be able to fully explain this phenomenon. 
More investigations should be done  especially with other experiments similar to MiniBooNE to investigate the impact of the pseudo-Hermitian physcis in the field of neutrino physics. 

\section*{Acknowledgments}

Thanks to you S. Bouzida for being in my life. Your support has been crucial to the achievement of this work. 


\end{document}